\documentstyle[epsfig]{mn}

\title[Deep HST imaging surveys and the formation of spheroidal galaxies]
{Deep HST imaging surveys and the formation of spheroidal galaxies}
\author[G. Rodighiero , A. Franceschini and G. Fasano.]
    {Giulia Rodighiero,$^1$\thanks{E-mail:rodighiero@pd.astro.it}
 Alberto Franceschini $^1$ and Giovanni Fasano $^2$\\
        $^1$Dipartimento di Astronomia di Padova, Vicolo dell'Osservatorio, 5, I-35122 Padova, ITALY\\
        $^2$Osservatorio Astronomico di Padova, Vicolo dell'Osservatorio, 5, I-35122 Padova, ITALY }

\begin{document}

\maketitle

\label{firstpage}

\begin{abstract}
We have extended our previous analysis of morphologically selected elliptical
and S0 galaxies in the Hubble Deep Field (HDF) North to include HST data in the
HDF South and the HDFS-NICMOS areas.  Our final sample amounts to 69 E/S0 
galaxies with $K<20.15$ over an area of 11 square arcmins. Although a moderately
small number over a modest sky area, this sample benefits of the best
imaging and photometric data available on high-redshift galaxies.
Multi-waveband photometry allows us to estimate with good accuracy the redshifts for
the majority of these galaxies which lack a spectroscopic measure.    
We confirm our previous findings that massive E/S0s tend to disappear
from flux-limited samples at $z>1.4$. This adds to the evidence that
the rest-frame colours and SEDs of the numerous objects found at $0.8<z<1.2$
are inconsistent with a very high redshift of formation for the bulk of stars,
while they are better consistent with protracted (either continuous or episodic)
star-formation down to $z \le 1$.
These results based on high-quality imaging on a small field can be complemented
with data from colour-selected Extremely Red Objects (EROs) on much larger sky areas: 
our claimed demise of E/S0s going from $z=1$ to $z=1.5$ is paralleled by a similarly
fast decrease in the areal density of EROs when the colour limit 
is changed from $(R-K)=5$ to $(R-K)=6$ (corresponding to $z\simeq 1$ and $z\simeq 1.3$ \
respectively).
Altogether, the redshift interval from 1 to 2 seems to correspond to a very
active phase for the assembly of massive E/S0 galaxies in the field, 
and also probably one where a substantial fraction of their stars are formed.

\end{abstract}


\section{Introduction}

The evolution with redshift of the mass function of bright galaxies is an 
important discriminant among competing cosmogonic scenarios. 
Present instrumentation, including HST and large optical telescopes
on ground, is close to provide decisive tests of these alternative schemes
for galaxy formation.

The surface density of Extremely Red Objects (EROs) has been recently
used by Daddi et al.~(2000) to constrain the epoch of formation and
the evolution of elliptical galaxies in the field. These authors
claim that the density of EROs with $(R-K) \ge 5$ (the typical colour
of an old elliptical galaxy at $z \ge 1$) is consistent with the
passive evolution of giant ellipticals, implying an early formation of
the spheroidal population through violent starbursts at high redshifts
($z \ge 2.5$). This formation paradigm for ellipticals has the nice
feature to make clear and testable predictions.

On the other side, hierarchical galaxy formation models (Kauffmann \&
Charlot 1998) predict that elliptical galaxies form through the
merging of disk galaxies at moderate redshifts. This scenario is supported by other 
studies emphasizing a lack of massive ellipticals at redshift greater
than $z \sim$~1.5 (Zepf~1997; Franceschini et al.~1998, hereafter FA98;
Barger et al.~1999).

In particular, FA98 studied the photometric and statistical properties
of a complete sample of early--type galaxies with $K < 20.15$ in the 
HDF North. 
The FA98 sample was selected on a deep $K$-band image from Kitt Peak, 
by applying a rigorous morphological classification
scheme based on the observed surface brightness profiles. 
FA98 found that the vast majority of bright early--type galaxies in the
HDF-N are located at $z\le 1.3$ and display colours indicative of a
fairly wide range of ages. There is a remarkable
absence in the FA98 sample of objects with $z > 1.3$. According to the
passive evolution scenario, these objects should be easily detected, 
due to the luminous star formation phase expected to
happen at these redshifts.

To enforce the statistical significance of these results, an extension
of the analysis to different sky regions would be critical. 
For this reason, we extend in this paper the study to the
HDF--South WFPC2 and NICMOS fields, by selecting a new complete sample 
of early--type galaxies and following the same prescriptions outlined 
in FA98 to process the data.

For consistency with previous analyses, we assume $H_0=50$ 
$Km~s^{-1}$ $Mpc^{-1}$ and $q_0=0.15$, with zero cosmological constant $\Lambda$. 

\section{Sample selection}

The present analysis makes use of a collection of high quality, deep
images, well suited for the study of field galaxies and covering three
different areas: (a) the HDF-North, (b) the HDF-S WFPC2 area and the
(c) HDF-S NICMOS field. Our galaxy catalog includes early-type
objects selected in the near infrared in such fields.

\subsection{Optical and near infrared imagery}
We defer the reader to FA98 for details on the HDF-N observations.
The WFPC2 strategy adopted for the HDF-S (Williams et al.~1998) 
was similar to that adopted for the HDF-N. 
In particular, the images were taken in the same four bands (F300W, 
F450W, F606W, F814W), with similar total exposure times.
For each filter, several exposures, obtained in
dithering mode, have been combined in a single image with pixel 
scale of $\sim 0.04^{\prime\prime}$. The sky area covered by both 
HDFN and HDFS is 10.6 arcmin$^2$.

Da Costa et al. (1998) observed the HDF-S WFPC2 area from optical to
near-IR at the ESO 3.5 New Technology Telescope (NTT). The optical
observations were carried out using the SUSI2 camera in binned mode,
yielding a scale of 0.16 arcsec/pixel.
Infrared observations were obtained in the $J$, $H$
and $K$ bands, using the SOFI camera (scale of 0.29 arcsec/pixel).

The HDF-S NICMOS images were acquired using 
camera 3 with a pixel size of 0$^{\prime\prime}$.2 and the filters F110W,
F160W and F222W. 
The dithered images were processed, drizzled and
combined giving a  final pixel size of 0$^{\prime\prime}$.075 and
covering an angular area of $\sim$1 arcmin$^2$.

The HDF-S NICMOS field has also been observed as part
of the VLT--UT1 Science Verification program, using the Test Camera
(VLTTC) at the Cassegrain focus and the $U$, $B$, $V$, $R$, $I$ filters. The
total field sampled covers 1.5$\times$1.5 arcmin$^2$ and the scale of
processed images is of 0.092 arcsec/pixel (Fontana et al. 1999). 

\subsection{The photometric filter}
The criteria adopted to ensure a careful selection of the sample in
the HDF-N, as well as the accurate treatment of photometric data in
seven passbands ($U$, $B$, $V$, $I$: from HST imaging, by Williams et al.~1996;
$J$, $H$, $K$: from KPNO imaging, Connolly et al.~1997) have been
discussed in FA98 and Rodighiero et al.~(2000). For the other fields
we have applied the same selection scheme used by FA98. We refer the
reader to those papers for a thorough description of the procedure, while
for convenience we summarize the main steps below.

\underline{\bf HDF-South-WFPC2 field}\\
Our primary selection is in the $K$-band, in order to minimize the
biases due to the effects of $K$ and evolutionary corrections. The
sample of galaxies has been extracted from the SOFI-$K$ image, running
SExtractor (Bertin $\&$ Arnouts 1996). As in FA98, to determine the
limit of completeness in the $K$ band ($K_L$) for inclusion in our
sample, we have used large simulated fields, including several
HDF-like toy galaxies and mimicking the image quality of the $K$-band
image. Running SExtractor on these synthetic frames allowed us to
check its performance, determining the biases and the standard
deviations of the magnitude estimates as a function of the surface
brightness. We then derived the corrections that must be applied
to the SExtractor photometry and the limit for completeness $K_L=20.2$
(in the standard system). From the simulations we derived also
an evaluation of
the $K$-band limiting surface brightness observable in the field, 
$\mu_K~ \sim 23 mag/arcsec^2$.
An analogous procedure was used to derive,
for each object in the selected sample, the corrected magnitudes in
the $J$ and $H$ bands. We did not correct optical HST magnitudes.

\underline{\bf HDF-South-NICMOS field}\\
The same procedure has been adopted for the HDF-S NICMOS field. In
this case, however, we have chosen the F160W-HST band ($\sim H$) 
for source selection, because of the higher noise
in the F222W image ($K$ band).
The importance of this field concerns its depth in the
near-IR (observed from the space), that allows to reach 
deeper limiting magnitudes. In particular, from the
simulations we derived a magnitude of 22.05 for the limit of
completeness in the $H$ band. As for the HDF-S-WFPC2 field,
we computed the corrections for the $J$ and $K$ bands, and also for the
ground based observations in the $U$, $B$, $V$, $R$ and $I$ bands.
Note that the selection in the F160W band does not introduce appreciable
bias with respect to the nominal $K$ band selection adopted in 
the other fields, because the typical spectra for $z\sim 1$ E/S0
galaxies are flat between 1.5 and 2.2 micron and are
contributed by similarly old stellar populations.


\begin{figure}
\begin{center}
 {\centering \leavevmode 
 \resizebox{9cm}{!}{\includegraphics{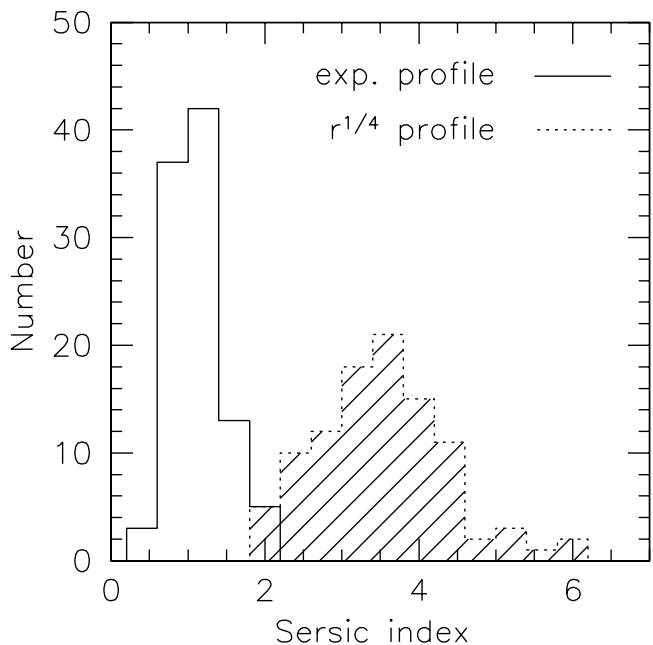}} }  
 \caption{Histograms of the Sersic index $n$ obtained by running GASPHOT on a
crowded artificial field containing 200 toy galaxies with different
magnitudes and radii. Galaxies with exponential profiles (100 objects)
are binned with the solid line histogram, whereas the dotted line
histogram refers to galaxies with $r^{1/4}$ profiles.}
 \label{figfitk}
\end{center}
\end{figure}

\begin{figure}
\begin{center}
 {\centering \leavevmode 
 \resizebox{9cm}{!}{\includegraphics{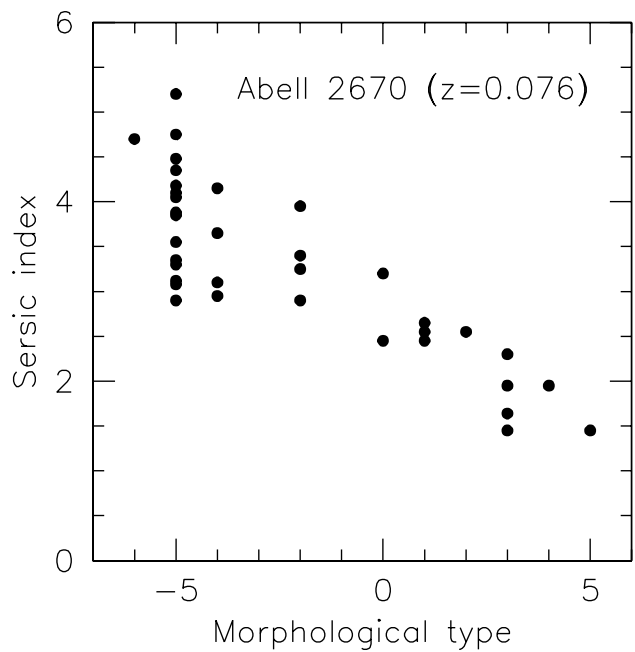}} }  
 \caption{Correlation between the Sersic index ($n$) and the de~Vaucouleurs
morphological type (T) for a sample of galaxies in the cluster
Abell~2670. The morphological types were derived by visual inspection
of the individual galaxies, complemented by luminosity and geometrical
profiles obtained from detailed surface photometry (Fasano et
al.~2001).}
 \label{figfitk}
\end{center}
\end{figure}

\subsection{The morphological filter}       

We have used the Galaxy Authomatic Surface PHOTometry tool (GASPHOT;
Pignatelli and Fasano 1999, 2000) to perform the morphological
analysis of all galaxies brighter than the proper limiting
IR-magnitudes in the WFPC2 and NICMOS fields of the HDF-S.

This tool operates on CCD images and consists of three different
steps: i) first, it exploits the basic extraction capabilities of
SExtractor to perform a simultaneous slicing of all selected objects
with a given step of surface brightness (we used $0^m.1$), providing
luminosity, ellipticity and position angle profiles for each object as
a function of the equivalent radius of the isophote; ii) secondly, it
gives an accurate, smooth representation of the $PSF$ by averaging
(with a $k$--sigma clipping procedure) the luminosity profiles of the
stars identified in the frame; iii) finally, it uses the generalized
de~Vaucouleurs law $\mu_r-\mu_0\propto (r/r_e)^{1/n}$ (where $\mu_r$
is the surface brightness at radius $r$ and $r_e$ is the half light
radius; Sersic~1968, see also Ciotti~1991), convolved with the proper
$PSF$, to fit the equivalent luminosity profiles of galaxies,
providing unbiased estimates of the profile parameters. The performances
of GASPHOT are presently undergoing a systematic investigation
(Pignatelli and Fasano 2001) by means of both extensive simulations of
toy galaxies of different morphological types and comparison with
galaxy samples whose morphology has been previously estimated by
visual inspection of images and profiles. Figures 1 and 2 illustrate
some preliminary results of this analysis. In particular, in Figure~1 we
report the histograms of the Sersic index $n$ obtained by running
GASPHOT on a crowded artificial field containing 200 toy galaxies
(with different magnitudes and radii) having exponential (100 objects)
or $r^{1/4}$ luminosity profiles. In Figure~2 we show the correlation
between the Sersic index ($n$) and the de~Vaucouleurs
morphological type (T) for a sample of galaxies in the cluster
Abell~2670. In this case the morphological types were derived by
visual inspection of the individual galaxies, complemented by
luminosity and geometrical profiles obtained from detailed surface
photometry (Fasano et al.~2001). From both the above figures
the value of $n$ given by GASPHOT appears to be a good indicator
of the `{\it true}' morphological type. 

According to Figure~1, after running GASPHOT on our high resolution
images, we decided to exclude from the final samples the galaxies with
$n\le$2 (considered late--type).  In addition, we rejected the
galaxies whose early-type classification was judged dubious at a
visual inspection (3 objects in the WFPC2 field and 1 object in the
NICMOS field). The final sample includes the 35 E/S0 galaxies
extracted from the HDF--N in FA98, 28 galaxies from the WFPC2 field
and 11 galaxies from the NICMOS field, for a total of 74 objects. Note
that only 6 galaxies in the NICMOS field are brighter than the
completeness limit for the WFPC2 fields ($K_L=20.15$).

In Table 1 we report the data of our sample: coordinates (at J2000),
photometric redshifts and the observed $K$ magnitudes
(the data for HDFN galaxies have been published in FA98).

\section{ANALYSIS OF THE BROAD BAND SPECTRA} 
\subsection{Tools for spectral synthesis} 
The observed UV-optical-near/IR spectra of the sample of E/S0 galaxies
presented in this work have been fitted with spectrophotometric models
describing their integrated emission. In order to check the
consistency of our approach, we decided to compare the
results from two different codes: GRASIL (Silva et al 1998) and PEGASE
(Fioc \& Rocca-Volmerange 1997). As in FA98, we have considered two
models describing the spectral behaviour of elliptical galaxies.  
We adopt a Salpeter IMF with a lower limit $M_l=0.15
M_{\odot}$ and a Schmidt-type law for the star formation rate (SFR):
$\Psi(t)=\nu M_g(t)$, where $\nu$ is the efficiency and $M_g(t)$ is
the residual mass of gas.  A further relevant parameter is the
timescale $t_{infall}$ for the infall of primordial gas. The evolution
patterns for the two models considered here are obtained with the
following choices of the parameters: 
Model 1: $t_{infall}=0.1~Gyr$, $\nu=2~Gyr^{-2}$;
Model 2: $t_{infall}=1~Gyr$, $\nu=1.3~Gyr^{-2}$. 
The corresponding SF laws have a maximum at
galactic ages of 0.3 (Mod 1) and 1.4 Gyr (Mod 2), and are truncated at
1 (Mod 1) and 3 Gyr (Mod 2) to mimic the onset of galactic winds.
 
A $\chi^2$ fitting procedure has been adopted for the
determination of the best solution for each galaxy spectrum. 
We use GRASIL as our reference model and use PEGASE only for
comparison (the two give equally good fits).

\subsection{Photometric redshifts and completeness}

An application of spectrophotometric models is the
determination of photometric redshifts. Our sample in the HDF-S lacks
a complete spectroscopic coverage (40\% of sample objects in the HDF-N
have measured redshift). We deemed useful to compare the GRASIL and PEGASE
redshift estimates. There is a slight tendency for GRASIL to 
predict larger values of $z$, but in general the agreement is 
quite good ($\sim 10\%$).

The distribution in redshift of a flux limited population
of galaxies provides a constraint on its evolutionary history
and on the epoch of formation. However, selection effects have to be
carefully taken into account, in particular the surface
brightness limit, the cosmological dimming and the K-correction.
Our selection in the $K$-band minimizes K- and evolutionary corrections
as a function of redshift. Our evaluation of the
effective radii allows us to control the effects of the limiting
surface brightness ($\mu_K \sim 23
mag/arcsec^2$) observable in the $K$-band image. We found
that the cutoff in surface brightness in the $K$ image does not
affect the selection above our limit of $K_L=20.15$, for 
redshifts up to $z\sim 2.5$.

\section{Results and discussion}

\subsection{Redshift distributions}

Figure 3 reports the observed redshift distribution of elliptical
galaxies in the HDF-S (dashed histogram), compared with that in the
HDF-N (continuous line). The two distributions clearly differ:
the HDF-N distribution shows a peak
at $z \sim 1$, while in the HDF-S the peak is shifted toward lower
redshifts ($z \sim 0.5$). These differences can be explained in terms 
of clustering effects in the limited volumes sampled.
It is worth noticing the global lack of bright spheroids 
in both fields at high redshifts. In particular, early--type galaxies seem 
to disappear at $z \ge 1.5$. Since this conclusion is now based on a
significantly larger area than used by FA98, it is becoming more unlikely
that this is due to galaxy clustering effects [see further arguments 
about this in Sect. 4.3].

\begin{figure}
\begin{center}
 \epsfig{file=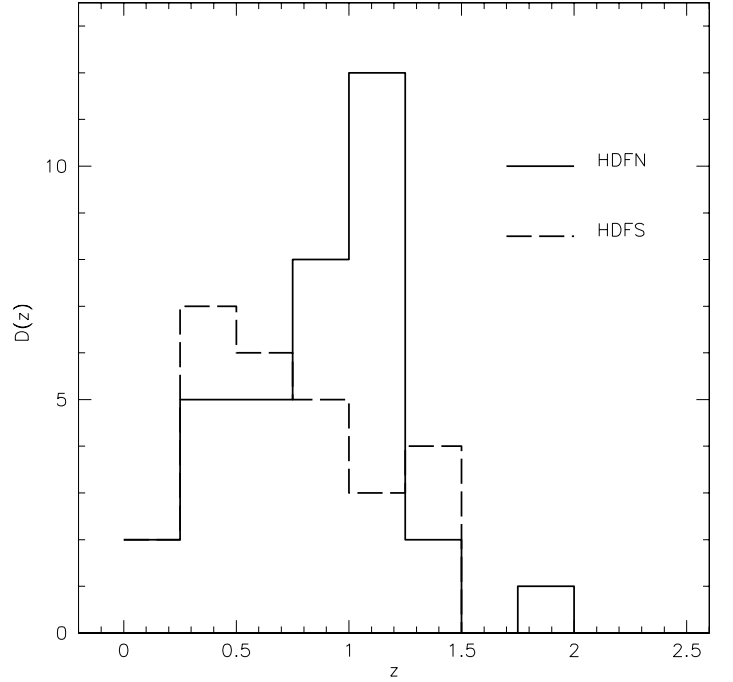,width=10cm}
 \caption{Observed redshift distribution in the HDFN (solid
line) compared with that in the HDFS (dashed line).}
 \label{figfitk}
\end{center}
\end{figure}

\begin{figure}
\begin{center}
 \epsfig{file=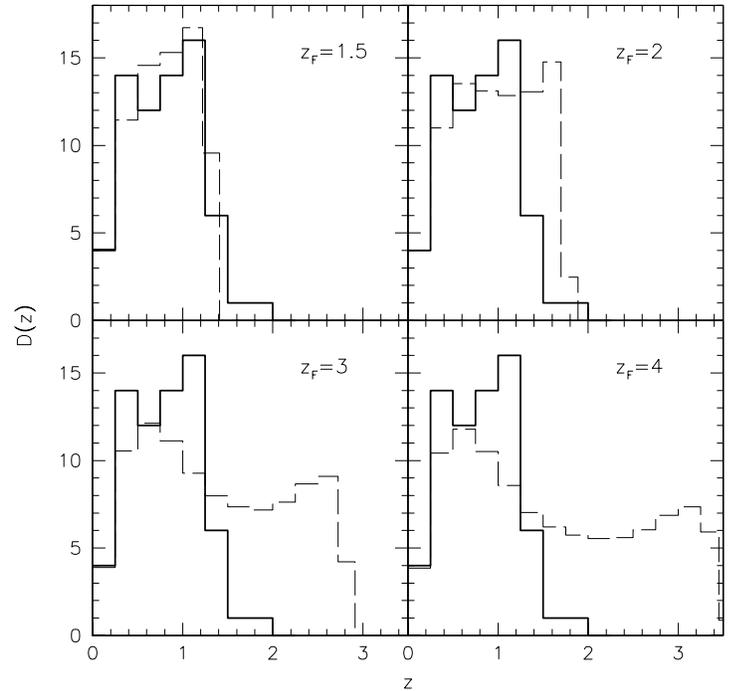,width=10cm}
\caption{Observed total redshift distribution (solid lines), including 
our sample objects from the HDFN, the HDFS and the NICMOS field 
at the limit magnitude of $K_L = 20.15$. The data are compared 
with predictions based on Model 1 for different redshift of 
formation $z_F$ (dashed lines); from top left to down right: 
$z_F$ = 1.5, 2, 3, 4.}
 \label{figfitk}
\end{center}
\end{figure}

In Figure~4 the observed distribution (continuous line) relative to
the total area of $\sim 11$ square arcminutes (HDFN + HDFS-WFPC2 +
NICMOS), for $K<20.15$, is compared with the corresponding
distributions predicted in the monolithic collapse scenario (dashed
lines). The four panels report the predictions of Model 1 for
increasing redshift of formation $z_F$. Low values of $z_F$ appear to be
consistent with the observed distribution ($z_F=1.5 \sim 2$), while an
early epoch of formation for field E/S0 seems to be ruled out
(see also Nakata et al. 1999). In particular, 39 E/S0 galaxies would be
expected at $z > 1.5$ for $z_F=3$ and 50 for $z_F=4$, compared with
the 2 observed.
Our data are more consistent with the hierarchical scenario, where 
large galaxies are predicted to form from the merging of smaller units. 
This finding is in agreement with Menanteau et al.~(2000), who 
estimate that at $z\sim 1$ about half the field ellipticals must be 
undergoing episodes of star-formation.

Among the solutions proposed to explain the demise of high-$z$ ellipticals,
FA98 consider the bias in the morphological filter against disturbed E/S0's
(consequence of merging), which may become significant at $z>1$. 
This point has been addressed by Rodighiero et al. (2001), who have analyzed,
in a way similar to that presented in this paper, a complementary 
sample of 52 late-type spiral and irregular galaxies selected in 
the same HDF-N area with total $K$-magnitudes brighter than K=20.47. One of
their main conclusions was about the fact that only 2 out of 52 galaxies 
in the sample have $z>1.4$. The lack of massive galaxies
of any morphological kinds at high $z$ supports that the
morphological selection of the spheroidal subsample is not
responsible for its cutoff at $z \ge 1.4$.

\subsection{Ages and baryonic masses for the sample galaxies}

The observed galaxy SEDs present two different spectral behaviours: 
red objects with spectra typical of old ellipticals, 
and blue objects that are flatter at all wavelengths, indicating the
presence of residual on-going star formation and a spread in the ages
of stellar populations.  Our 
$\chi^2$ fitting procedure lead us to the same conclusions as found by FA98: 
model 2 with protracted star-formation seems to better reproduce all
kinds of spectral behaviour of galaxies in the HDF-S,  providing a formally acceptable 
total $\chi^2$ of $\sim 290$ (7 band data for 28 HDF-S galaxies), corresponding to 
a reduced chi-square of $\chi^2_{\nu} \sim 1.5$.  Model 1 gives
acceptable fits only for galaxies with red spectra, while the total $\chi^2 \sim
400$ is unacceptably high ($\chi^2_{\nu} \sim 2$). 
This result supports the hypothesis that, on average, star formation activity
in these galaxies occurs over protracted periods (3 Gyr in the model), rather
than in short-lived single starburst (at least when considering our limited
set of models).

\begin{figure}
\begin{center}
 \epsfig{file=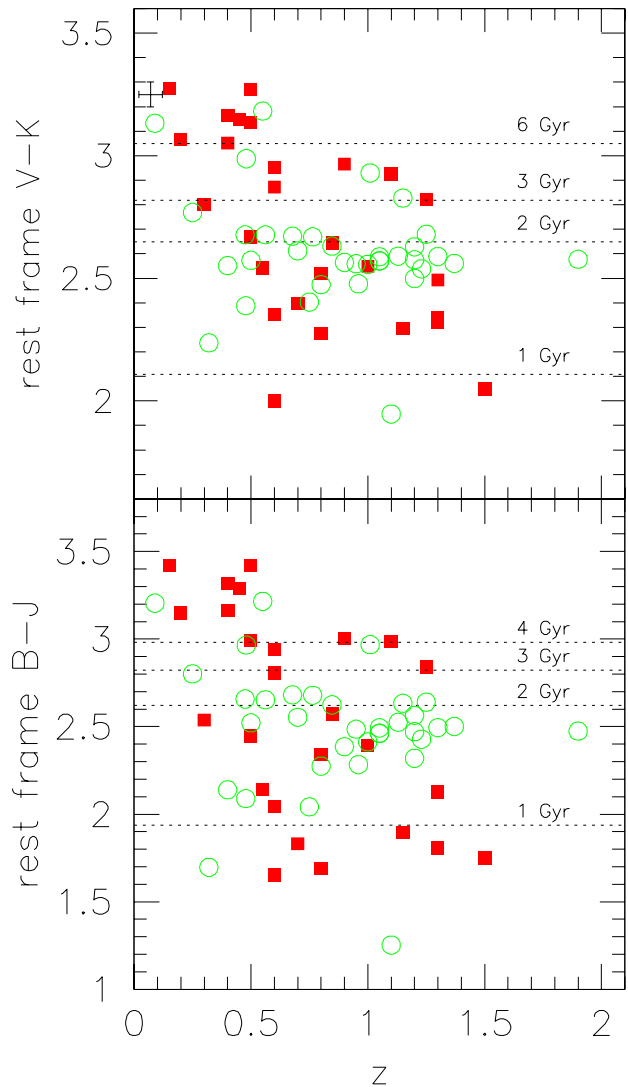,width=9cm}
\caption{Rest frame $(V-K)$ (upper panel) and $(B-J)$ (lower panel)
colours of early type field galaxies, compared with predicted
values for single stellar populations with solar metallicity.
Filled squares refer to HDFS objects, open circles to the HDFN.}
 \label{figfitk}
\end{center}
\end{figure}

We compare in Figure~5 the rest-frame $(V-K)$ and $(B-J)$ colours 
(computed from best-fitting solutions) of 
early--type galaxies as a function of redshift
with the predicted colours of single stellar populations with
solar metallicity. The colours for HDF-N ellipticals (open circles)
are consistent with those found for HDF-S objects (filled squares),
whereas the large spreads observed in both colour distributions
correspond to a wide range of ages for the galaxies in the sample,
typically from 1 to 5 Gyr. The figure shows a tendency for high
redshift ellipticals to be intrinsically bluer, 
in particular in the HDF-S, while the clumped distribution observed 
in the HDF-N at $z\sim 1$ and $V-K \simeq 2.5$  likely indicates 
the presence of clustering at this redshift.

We plot in Figure~6 the baryonic masses of our sample galaxies
as a function of redshift, and compare them with
predictions of model 1 (dotted lines) and model 2
(solid lines) corresponding to the lowest galaxy mass still detectable within the
flux limit of the survey ($K_L=20.15$) as a function of redshift. Different
curves in the figure refer to various redshifts of formation $z_F$
(numbers attached to each curve).

\begin{figure}
\begin{center}
 \epsfig{file=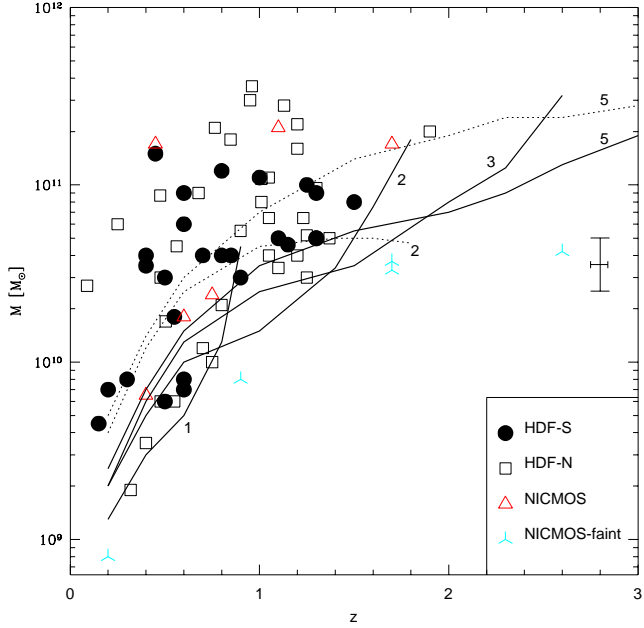,width=9cm}
\caption{The observed baryonic mass of the sample galaxies at $K_L=20.15$
is plotted against the redshift. The curves represent the
predictions of model 1 (dotted lines) and model 2
(solid lines) for the lowest galaxy mass still detectable within the
flux limit of the survey ($K_L=20.15$) as a function of redshift
for various redshift of formation (number attached to each curve).
The error bar shows the mean uncertainty on the determination of
the mass from spectral fitting, when considering the two different
models.} 
 \label{figfitk}
\end{center}
\end{figure}

In agreement with the results of Figure 4, only low values of 
$z_F$ ($\le 2$) are consistent with the observed mass distribution.  
In particular, the predictions of model 1 with $z_F=5$ 
are inconsistent with the observed distribution,  half 
of the sample galaxies having estimated masses lower than
the prediction. This model fails to reproduce
the broad-band spectra of each single galaxy.
Model 2 with $z_F=2$ seems to better reproduce the observations,
and provides an appropriate match with the lower envelope
of the mass distribution. Several galaxies at $z<1$ have masses
requiring a lower redshift of formation ($z_F=1$).

We stress that
Figure 6 does not imply that ellipticals at $z \ge 1.5$ do not exist.
Rather, according to the prescriptions of hierarchical models 
(Kauffmann \& Charlot 1998), they are likely to be missed in our 
flux limited sample due to their lower mass (and luminosity).
This is strongly suggested by the location in Figure~6 of the five
objects selected in the NICMOS field and fainter than $K=20.15$ 
(three-pointed stars). 

\begin{figure*}
\begin{center}
 \epsfig{file=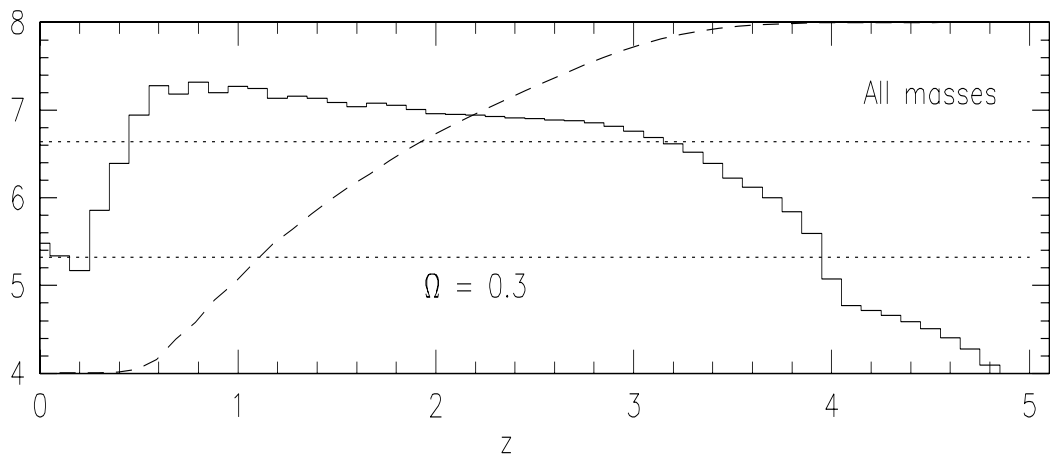,width=16cm,height=8cm}
\caption{Distributions of the density of stellar mass per unit
comoving volume generated in the redshift bin as a function of
redshift (in $M_{\odot}~Mpc^{-3}$ born in the redshift bin).
The dashed lines represent the cumulative
distributions of the mass density generated as a function of
redshift, on a linear scale ranging from $0\%$ to $100\%$. The
dotted horizontal lines mark the 33 and 66 percentiles.}
 \label{massbin}
\end{center}
\end{figure*}

Benitez et al. (1999) have detected and identified in the NICMOS
field 4 early-type galaxies at $z>1$ with $H(AB)<25$ (corresponding
to $H_{ST} \sim 23.6$). 
Of these, ET2 (in the Benitez et al. list)
is fainter than our limit of completeness, ET1 and ET3 are included 
in our sample, while ET4 has been rejected as it is 
undetected in the VLT images and SEDs based on the NICMOS near-IR data
cannot be satisfactory fitted with any of our galaxy models.

Benitez et al. suppose that the presence of luminous
red galaxies at high $z$ in a small field indicates that 
evolved ($z_F \ge 5$), massive galaxies are common at $z>1$.
They suggest that the absence of distant ellipticals claimed 
by others (FA98, Menanteau et al. 1999, Barger et al. 1999) may 
be due to the depth of the NICMOS data compared with ground-based 
IR data (about 3 magnitudes) or to clustering effects.

The present study is not inconsistent with the results reported by
Benitez et al.. First, we confirm the presence of a substantial number (21 out of
74) of E/S0 galaxies at $z$ between 1 and 1.4. We also prove the existence
of E/S0 galaxies above $z=1.5$. In particular if we include the fainter
NICMOS data, we find 4 early-types in NICMOS at $1.5<z<3$ 
with $H<22.05$.
Our claim is rather that the baryonic masses found for these high-$z$
objects (typically few $10^{10}\ M_\odot$ units) are typically lower than 
expected on consideration of the very massive ($M>10^{11}\ M_\odot$) galaxies 
detected around $z=1$ in our areas.
The areal density of massive E/S0 as a function of redshift shows a drastic
decrease, not a complete disappearance, above $z\sim 1.4$.

All this seems to favour an evolutionary scenario where the bulk of 
luminous spheroids are assembled at relative recent cosmic epochs 
($z_F \le 2$), although not as recent as supported by
Kauffmann \& Charlot (1998). 

Obviously, the assembly of stellar populations into the
dynamically relaxed shapes of spheroids could have occurred
later than the birth epochs of stars. The distribution of
redshift of formation for stellar populations in 
E/S0 has been quantified by FA98 in their HDF-N analysis.
This showed that the bulk of stars formed at cosmic epochs
between $z=1$ and $z=3$.

The results of a similar analysis of our combined sample is 
reported in Figure~7. The histogram in the Figure shows
the mass in stars formed per unit comoving volume in the
redshift bins. This estimate is based on our best-fit 
photometric model solutions for all galaxies in the sample,
by adding up the individual rates of star formation in the redshift
bins. The mass is corrected for the portion of the
luminosity function not sampled by our $K$-flux
selection using the local observed LF in the $K$-band
(see FA98 for details on the computation).

As shown by the figure, the history of formation
of stars for this combined HDFN, HDFS, NICMOS sample
is different from the partial result
based on the HDF-N. The dashed curve is the normalized
integral distribution: in the present case and for our
adopted cosmology , $30\%$ of stars are formed at at $z<1$, $40\%$ between
$z=1$ and 2, and the remaining $30\%$ at $z>2$.

\subsection{Effects of cosmic variance on searches for high-z spheroids}

Clearly, cosmic variance affects results based on small areas 
such as the HDFs (see Sect. 4.1 and Figure 3). 
Daddi et al.~(2000; hereafter DA) showed that the surface density
variations of the Extremely Red Objects (EROs) population 
found in their survey can contribute to explain
the discrepant results obtained by different authors about the density
of $z>1$ ellipticals. Similarly, Eisenhardt et al. (2000), using a
selection of $z>1$ elliptical galaxies based on the $(J-K)$ colour,
examined the surface density variation between their $K$-selected survey 
(124 $arcmin^2$) and the HDF-N and argued that clustering can affect
results based on small area surveys. Actually, the cosmic variance 
is clearly apparent in our HDF-S and HDF-N catalogues (see Figure 3, 
see also Lee \& Hwang 2000). 

Some insights about the effects of cosmic variance on our conclusions can be 
obtained by comparing results based on our small ($11$ $arcmin^2$) area
with data on the surface density 
of EROs derived by DA for a much larger sky area ($\sim 700~arcmin^{2}$). 
DA define as EROs those objects with colours $(R-K) \ge 5$. 
They could be passively evolving galaxies at $z >1$ (with red colours because
of the large K-correction) or heavily dust-reddened, star-forming
galaxies or AGN.  Using the above definition of EROs, DA found a
surface density of 0.67 EROs $arcmin^{-2}$ at the limit of their
survey ($K\le 19.2$). A more stringent colour criterion of $(R-K) \ge 6$
should select objects at $z \ge 1.3$. At the limit of $(R-K) = 6$
they found a surface
density of only 0.10 EROs $arcmin^{-2}$. If we adopt the same magnitude
limit of $K\le 19.2$, the corresponding surface densities for
$z>1$ and $z>1.3$ ellipticals that we find in our sample are: 
0.88 $arcmin^{-2}$ ($R-K \ge 5$) and 0.17 $arcmin^{-2}$ ($R-K \ge 6$). 
Thus, {\sl in spite of the very different survey areas and selection criteria, 
the EROs statistics and our results about the fraction of E/S0 galaxies 
at high-z turn out to be in quite good agreement.
This supports our conclusions about the demise of bright massive 
spheroids at high redshifts.}

This fast decrease of the areal density of E/S0 at $z>1.5$ is likely to
correspond to a fast evolution of the mass function at these epochs.
Altogether, a very active phase not only for the assembly of massive E/S0 
galaxies in the field but also for the origin of stellar population
seems to happen around $z\sim 1$ to 2.

\section{CONCLUSIONS}

We have analyzed a $K$-band flux-limited sample of 69 E/S0 galaxies
in three different areas (HDF-North, HDF-South and the NICMOS field, for a total
of $\sim 11$ arcmin$^2$), where deep HST data are available.
This sample is uniquely suited, as for the quality of the imaging and photometry,
to the study of high-redshift spheroidal galaxies.

Our main conclusions can be summarized as follows.
\begin{itemize}

\item
We confirm the finding by FA98 that massive E/S0s tend
to disappear from flux limited samples at $z>1.4$.
At least the assembly of massive spheroids has to take place efficiently
in the redshift interval $1<z<2$.

\item
The large spreads in the rest-frame colour distributions 
(both $V-K$ and $B-J$) for the early-type galaxies observed 
at $z\sim 1$ in our sample indicate a wide range of ages, 
typically from 1 to 5 Gyr. These
objects are inconsistent with a very high redshift
of formation for the bulk of stars, and are consistent with 
protracted star-formation (either continuous or episodic) down to $z \le 1$.

\item
Averaged over our (small) selected area, the bulk of stars (40\%) are formed in the 
redshift interval $1<z<2$. Of the remaining fraction, half form at $z<1$ and
half at $z>2$.

\item
Our finding of a demise of E/S0s at $z>1.5$ in small fields
is confirmed on much larger areas by a similarly fast decrease in the number
density of EROs, when the colour limits are changed from
$(R-K)>5$ to $(R-K)>6$ (corresponding to an increase of the redshift cutoff
from $z\simeq 1$ to $z\simeq 1.3$, for typical evolved stellar spectra).

\item
Hierarchical models for the formation of spheroidal galaxies may naturally
explain the present results.

\end{itemize}

\begin{table*}
\caption{Data on the sample galaxies}
\begin{tabular}{|l|c|c|c|c|c|c|c|c|}
\hline
\hline
id&~&$R.A.$&~&~&$DEC.$&~&$z$&$K$\\
\hline
~&$h$&$m$&$s$&$^o$&$'$&$''$&~&~\\
\hline                                                          
HDFS1  & 22  &  32 &  54.08 &  -60 &   31 &  42.75 & 0.50 &    19.73\\
HDFS2  & 22  &  32 &  52.23 &  -60 &   31 &  52.80 & 1.30 &    19.85\\
HDFS3  & 22  &  32 &  53.37 &  -60 &   32 &   1.27 & 1.30 &    19.59\\
HDFS4  & 22  &  32 &  50.28 &  -60 &   32 &   3.30 & 0.60 &    19.44\\
HDFS5  & 22  &  32 &  55.72 &  -60 &   32 &  11.46 & 0.60 &    18.16\\
HDFS6  & 22  &  32 &  54.78 &  -60 &   32 &  15.46 & 0.50 &    18.27\\
HDFS7  & 22  &  32 &  57.75 &  -60 &   32 &  33.01 & 0.55 &    19.23\\
HDFS8  & 22  &  32 &  50.90 &  -60 &   32 &  43.02 & 0.40 &    17.69\\
HDFS9  & 22  &  32 &  54.05 &  -60 &   32 &  51.67 & 0.60 &    19.86\\
HDFS10 & 22  &  32 &  51.66 &  -60 &   33 &   6.03 & 1.10 &    19.72\\
HDFS11 & 22  &  32 &  53.92 &  -60 &   33 &  13.40 & 0.20 &    18.28\\
HDFS12 & 22  &  33 &   1.88 &  -60 &   33 &  16.36 & 0.30 &    18.89\\
HDFS13 & 22  &  33 &   2.75 &  -60 &   33 &  22.05 & 0.45 &    16.83\\
HDFS14 & 22  &  32 &  57.04 &  -60 &   33 &  23.01 & 1.15 &    19.62\\
HDFS15 & 22  &  32 &  52.16 &  -60 &   33 &  23.88 & 1.30 &    19.91\\
HDFS16 & 22  &  32 &  53.02 &  -60 &   33 &  28.49 & 1.00 &    18.44\\
HDFS17 & 22  &  32 &  54.99 &  -60 &   33 &  29.03 & 1.25 &    19.22\\
HDFS18 & 22  &  32 &  57.09 &  -60 &   33 &  28.91 & 0.90 &    19.93\\
HDFS19 & 22  &  32 &  45.32 &  -60 &   33 &  32.51 & 0.15 &    18.44\\
HDFS20 & 22  &  32 &  52.35 &  -60 &   33 &  33.05 & 0.80 &    20.10\\
HDFS21 & 22  &  32 &  51.50 &  -60 &   33 &  37.59 & 0.60 &    18.46\\
HDFS22 & 22  &  33 &   5.10 &  -60 &   33 &  46.16 & 1.90 &    19.86\\
HDFS23 & 22  &  32 &  58.60 &  -60 &   33 &  46.56 & 1.60 &    19.56\\
HDFS24 & 22  &  32 &  46.89 &  -60 &   33 &  54.83 & 0.40 &    18.05\\
HDFS25 & 22  &  32 &  52.24 &  -60 &   34 &   2.76 & 0.80 &    19.03\\
HDFS26 & 23  &  32 &  50.95 &  -60 &   34 &   5.00 & 1.50 &    19.21\\
HDFS27 & 22  &  32 &  48.90 &  -60 &   34 &   4.79 & 0.85 &    19.14\\
HDFS28 & 22  &  32 &  56.08 &  -60 &   34 &  14.20 & 0.70 &    19.59\\
HDFS29 & 22  &  33 &   0.48 &  -60 &   34 &  17.67 & 0.50 &    19.79\\
\hline
NICMOS1   & 22  &  32 &  51.10 &  -60 &   39 &   9.79 & 1.70 &    19.46\\
NICMOS2   & 22  &  32 &  50.69 &  -60 &   39 &   9.05 & 2.60 &    21.42\\
NICMOS3   & 22  &  32 &  53.41 &  -60 &   39 &   1.63 & 0.60 &    19.36\\
NICMOS4   & 22  &  32 &  53.70 &  -60 &   39 &   4.38 & 1.70 &    20.69\\
NICMOS5   & 22  &  32 &  52.13 &  -60 &   39 &   3.88 & 1.70 &    20.89\\
NICMOS6   & 22  &  32 &  53.02 &  -60 &   38 &  54.81 & 0.75 &    19.38\\
NICMOS7   & 22  &  32 &  52.86 &  -60 &   38 &  37.56 & 0.45 &    17.53\\
NICMOS8   & 22  &  32 &  51.11 &  -60 &   38 &  40.16 & 0.20 &    20.71\\
NICMOS9   & 22  &  32 &  55.44 &  -60 &   38 &  32.49 & 1.50 &    18.87\\
NICMOS10   & 22  &  32 &  50.69 &  -60 &   38 &  30.27 & 1.30 &    20.37\\
NICMOS11  & 22  &  32 &  53.43 &  -60 &   38 &  20.62 & 0.40 &    19.74\\
NICMOS12  & 22  &  32 &  52.05 &  -60 &   38 &  18.07 & 0.90 &    20.65\\
\hline
\hline
\end{tabular}
\end{table*}


\begin{thebibliography}{}

\bibitem[1999]{barger}
 Barger A.J., Cowie L.L., Trentham N., Fulton E., Hu E. M., Songaila A., 
 Hall D., 1999, AJ, 117, 102

\bibitem[1999]{benitez}
 Benitez N., Broadhurst T., Bouwens R., Silk J., Rosati P., 1999, ApJ, 515, L65

\bibitem[1996]{bertin}
 Bertin E., Arnouts S., 1996, A\&AS, 117, 393

\bibitem[1991]{ciotti}
 Ciotti L., 1991, A\&A, 249, 99

\bibitem[1997]{connolly}
 Connolly A. J., Szalay A. S., Dickinson M., Subbarao M. U., Brunner R. J.,
 1997, ApJ, 486, L11

\bibitem[1998]{dacosta}
 Da Costa L., et al., 1998, astro-ph/9812105

\bibitem[2000]{daddi}
 Daddi E., Cimatti A., Pozzetti L., Hoekstra H., Rottgering H.J.A.,
Renzini A., Zamorani G., Mannucci F., 2000, A\&A, 361, 535 (DA)


\bibitem[2000]{eisenhardt}
 Eisenhardt P., Elston R., Stanford S.A., Dickinson M., Spinrad H., Stern D., Dey A., 2000, astro-ph/0002468

\bibitem[2001]{fasano}
Fasano, G., Bettoni, D., D'Onofrio, M., Kj\ae rgaard, P., Moles, M., 2001, A\&AS, submitted.

\bibitem[1997]{fioc}
 Fioc M., Rocca-Volmerange B., 1997, A\&A, 326, 950

\bibitem[1999]{fontana}
 Fontana A., et al., 1999, AJ, 343, L19

\bibitem[1998]{franceschini}
 Franceschini A., Silva L., Fasano G., Granato G.L., Bressan A., Arnouts S., Danese L., 1998, ApJ, 506, 600

\bibitem[1998]{kauffmann}
 Kauffmann G., Charlot S., 1998, MNRAS 297, L23

\bibitem[2000]{lee}
 Lee M.G., Hwang N., 2000, astro-ph/0008286

\bibitem[2000]{menanteaub}
 Menanteau F., Ellis R.S., Abraham R.G., Barger A.J., Cowie L.L., 1999, MNRAS, 309, 208

\bibitem[2000]{menanteau}
 Menanteau F., Abraham R.G., Ellis R.S., 2000, astro-ph/0007114

\bibitem[1999]{nakata}
 Nakata F., Shimasaku K., Doi M., Kashikawa N. et al, 1999, MNRAS, 309, L25

\bibitem[1999]{pignatelli}
Pignatelli, E., Fasano, G., 1999, Ap\&SS 269, 657

\bibitem[2000]{pignatelli}
 Pignatelli E., Fasano G., proceedings of the Granada Euroconference "The
       Evolution of Galaxies.I-Observational Clues", astro-ph/0009037

\bibitem[2001]{pignatelli}
Pignatelli, E., Fasano, G., 2001, in preparation

\bibitem[2000]{rodighiero}
 Rodighiero G., Granato G.L., Franceschini A., Fasano G., Silva L., 2000,
 A\&A in press, astro-ph/0010131

\bibitem[1968]{sersic}
 Sersic J.L., 1968, Atlas de galaxias australes, Observatorio Astronomico, Cordoba

\bibitem[1998]{silva}
 Silva L., Granato G.L., Bressan A., Danese L., 1998, ApJ, 509, 103


\bibitem[1996]{williams}
 Willians R.E., et al., 1996, AJ, 112, 1335

\bibitem[1998]{williamsb}
 Willians R.E., et al., 1998, AAS, 193, 75.01

\bibitem[1997]{zepf}
 Zepf S., 1997, Nature, 390, 377


\end{thebibliography}
\end{document}